\documentclass[aps,prl,reprint,groupedaddress]{revtex4-2}

\usepackage{graphicx}% Include figure files
\usepackage{dcolumn}% Align table columns on decimal point
\usepackage{bm}% bold math
\usepackage{amsmath}
\usepackage{hyperref}
\usepackage{natbib}
\usepackage{subcaption}
\begin{document}

%\preprint{APS/123-QED}

\title{Probing Cosmic Ray Composition and Muon-philic Dark Matter via Muon Tomography}% Force line breaks with \\

\author{Cheng-en Liu}
\email{liuchengen@stu.pku.edu.cn}
\author{Rongfeng Zhang}%
\author{Zijian Wang}%
\author{Andrew Michael Levin}
\author{Leyun Gao}%
\author{Jinning Li}%
\author{Minxiao Fan}%
\author{Youpeng Wu}%
\author{Zibo Qin}
\author{Yong Ban}
\author{Zaihong Yang}
\author{Qite Li}
\email{Contact author: liqt@pku.edu.cn}
\author{Chen Zhou}
\email{Contact author: czhouphy@pku.edu.cn}%
\author{Qiang Li}
\email{Contact author: qliphy0@pku.edu.cn}%
\affiliation{%
 State Key Laboratory of Nuclear Physics and Technology,School of Physics,Peking University\\
 Beijing, 100871, China 
}%

\collaboration{PKMu Collaboration}%\noaffiliation

\begin{abstract}
This work presents a novel cosmic-ray scattering experiment employing a Resistive Plate Chambers (RPC) muon tomography system. By introducing the scattering angle between incident and outgoing cosmic-ray tracks as a key observable, this approach enables simultaneous studies of secondary cosmic-ray composition and searching for new physics. During a 63-day campaign, 1.18 million cosmic ray scattering events were recorded and analyzed. By performing combined template fits to the observed angular distribution, particle abundances are measured -- for example, resolving the electron component at $\sim 2\%$ precision. Furthermore, constraints are established on elastic muon–dark matter (DM) scattering cross-sections for muon-philic dark matter. At the 95\% confidence level, the limit reaches 1.61 $\times$ $10^{-17}$ $\rm{cm}^{2}$ for 1 GeV slow DM, demonstrating sensitivity limit to light muon-coupled slow DM, in scenarios where a strongly interacting dark matter component is captured and thermalized within the Earth, leading to large surface densities.

\end{abstract}

%\keywords{Suggested keywords}%Use showkeys class option if keyword
                              %display desired
\maketitle

%\tableofcontents

\section{Introduction}
Cosmic rays are high-energy charged particles originating from outer space, primarily composed of protons, helium nuclei, and a small fraction of heavier elements. Upon entering the Earth's atmosphere, these primary cosmic rays interact with atmospheric nuclei, generating cascades of secondary particles, such as muons, electrons, photons, and neutrons, collectively known as extensive air showers (EAS) \citep{rao1998extensive, grieder2001cosmic}. A portion of these secondary particles reach the Earth's surface and constitute a significant component of the natural background radiation at sea level.

Recent studies on secondary cosmic rays at sea level primarily focus on the muonic component, which is the most abundant \citep{BONOMI2020103768, Workman:2022ynf, https://doi.org/10.1029/92JA02672, PhysRevLett.83.4241, HAINO200435, ACHARD200415, ALLKOFER1971425, B_C_Rastin_1984, C_A_Ayre_1975, PhysRevD.19.1368}. While measurements of other particles, such as electrons and photons, were conducted over five decades ago, the limitations in detector technologies at that time led to significant uncertainties, ranging from 10\% to 20\% \citep{doi:10.1139/p68-411, grieder2001cosmic, Allkofer1970}. Over the past two decades, experimental data on these components have been limited, leaving the particle composition at sea level inadequately constrained. Accurate determination of the relative abundances of secondary cosmic-ray particles is crucial for understanding the energy spectrum and interaction mechanisms of primary cosmic rays, as well as for evaluating the effects of environmental factors such as solar activity, geomagnetic modulation, and atmospheric conditions \citep{miroshnichenko2015solar, sikdar_2023}. These measurements also play a vital role in testing particle interaction models, assessing ground-level radiation exposure, and developing cosmic ray–based techniques such as muon imaging \cite{BONOMI2020103768}. Particularly in searches using cosmic-ray muons as probes for dark matter (DM) \cite{PhysRevD.110.016017}, a precise accounting of secondary particle backgrounds is crucial for effective background suppression.

Muons provide a powerful link between fundamental research and practical applications. Both cosmic-ray muons and those generated at accelerators can be utilized to probe physics beyond the Standard Model (SM). Despite the remarkable success of the SM, it does not address several fundamental questions, such as the origin of neutrino masses and the nature of DM. In the absence of conclusive evidence for heavy new particles predicted by various extensions of the SM, attention has increasingly shifted toward exploring new scenarios involving light DM candidates with masses less than around 10 GeV/$c^2$, including those with muon-specific interactions \citep{essig2023snowmass2021cosmicfrontierlandscape, harris2022snowmasswhitepapernew, Bai2014, ParticleDataGroup:2024cfk}.

Muon scattering experiments have remained relatively rare to date. Early studies from the 1960s to the 1980s primarily focused on nuclear structure \cite{annurev:/content/journals/10.1146/annurev.ns.33.120183.002123}. More recent scattering experiments, such as NA64$\mu$ \cite{PhysRevLett.132.211803} and MUonE \cite{CARLONICALAME2015325}, have regained attention due to their potential to test the SM and search for new physics. In this context, the PKMu project at Peking University introduces a novel approach to DM detection based on cosmic-ray muon scattering \citep{PhysRevD.110.016017, doi:10.1142/S0217732325300083}. In contrast to accelerator-based active experiments, this work relies on passive cosmic-ray observations, utilizing the naturally occurring cosmic rays. If DM interacts preferentially with muons, it may deflect their trajectories in a detectable manner. By measuring the angular distribution of muon scattering events, it becomes possible to search for signs of slow-moving, light, muon-philic dark matter \cite{PhysRevLett.131.011005, PhysRevD.109.075027, PhysRevD.103.115031}.

In this study, a muon scattering detection system based on Resistive Plate Chambers (RPCs) \citep{LI201222, Li_2013} is constructed. Using this system, scattering measurements of cosmic-ray muons under ambient conditions are performed. From the angular distribution of muon deflections, the composition of secondary cosmic rays at sea level (corresponding to an altitude of approximately 44 meters above sea level, as in Beijing) is extracted. Furthermore, this method is used to place constraints on the muon-DM scattering cross section, demonstrating the feasibility of this approach for future DM searches.

\section{Experimental Setup}
The muon tomography system employs large-area glass RPCs featuring submillimeter spatial resolution -- a technology developed by the authors in 2012 \cite{PhysRevD.110.016017, LI201222, Li_2013} using LC delay-line readout. This readout scheme routes signals collected by each strip through a delay-line network, where the hit position along one dimension is determined by the time difference between signal arrivals at both ends of the line. With two orthogonally oriented readout strips, a single RPC detector achieves two-dimensional position reconstruction.

\begin{figure}[htbp]
    \centering
    \begin{subfigure}[b]{0.24\textwidth}%
        \includegraphics[width=\linewidth]{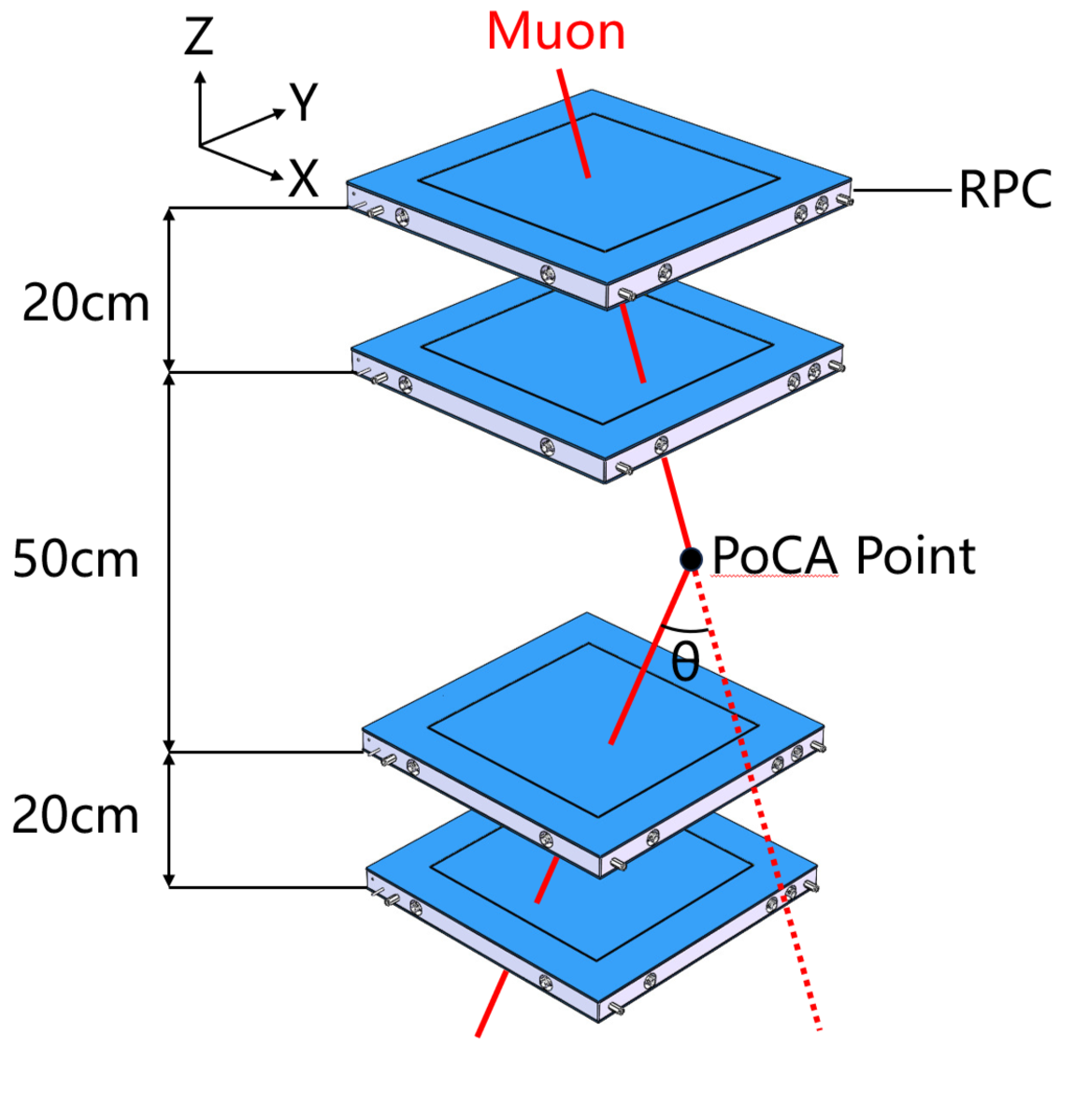}%
        \caption{}
        \label{fig:muon_tomography_system}
    \end{subfigure}%
    \hfill
    \begin{subfigure}[b]{0.24\textwidth}%
        \includegraphics[width=\linewidth]{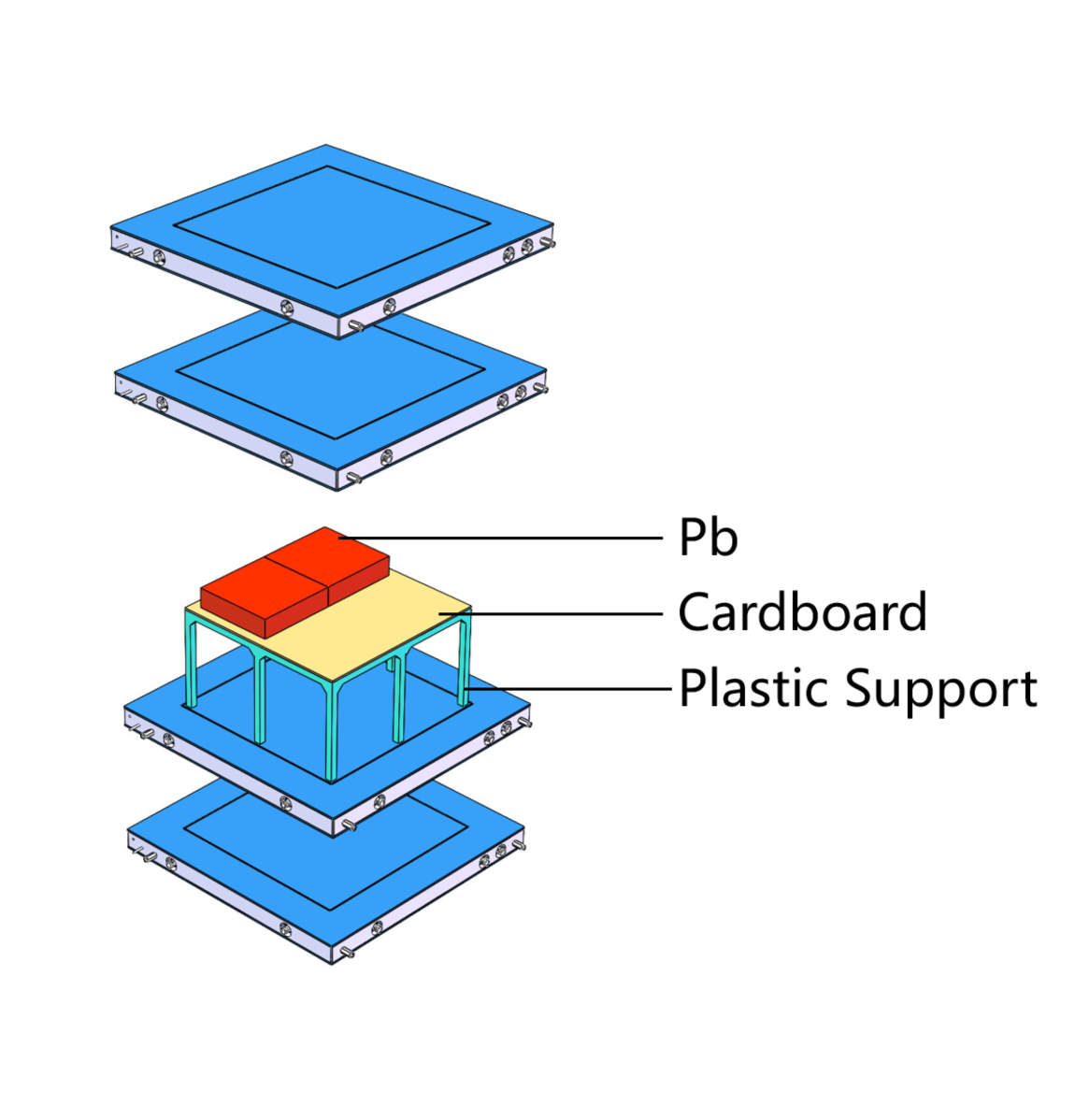}%
        \caption{}
        \label{fig:pb_experiment_setup}
    \end{subfigure}%
    \captionsetup{justification=raggedright} % 局部左对齐
    \caption{Experimental configurations. (a) Physics RUN: Four vertically aligned glass RPCs (28 $\times$ 28 $\rm{cm}^2$) with spacings of 20, 50, and 20 cm; the muon scattering angle $\theta$ is defined. (b) Control RUN: Two lead blocks (12 $\times$ 12 $\times$ 3 $\rm{cm^3}$) placed 150 mm above the third RPC, centered at the origin ($X$ $\in$ [-120, 0] mm, $Y$ $\in$ [-120, 120] mm).}
    \label{fig:setup}
\end{figure}

As shown in Fig.~\ref{fig:muon_tomography_system}, four RPC detectors are arranged vertically ($Z$-direction) with spacings of 20 cm, 50 cm, and 20 cm, forming the muon tomography system. Each RPC detector provides a spatial resolution of 0.7 mm and a two-dimensional position reconstruction efficiency of $\sim$ 85\% over its 28 $\times$ 28 $\rm{cm}^2$ active area.

Two dedicated runs were conducted: Physics Run: 63-day continuous data acquisition under laboratory air conditions (non-vacuum), yielding 1.18 million effective events (triggered by coincident T-signals from top/bottom RPC pairs with valid 2D positions in all layers). Control Run: 11-day background validation test with a lead block target (two 12 $\times$ 12 $\times$ 3 $\rm{cm}^{3}$ blocks coverage: $X$ $\in$ [-120, 0] mm, $Y$ $\in$ [-120, 120] mm, positioned 150 mm above the third RPC) (Fig.~\ref{fig:pb_experiment_setup}).

Events were filtered to include only those with validated 2D positions across all four RPC layers. Data collection began on February 12, 2025, with the Physics Run providing the primary dataset for Cosmic Ray Composition test and dark matter searches, while the Control Run constrained background models.

\section{Simulation}

The simulation framework combines CRY (Cosmic-ray Shower Generator) \cite{4437209} for cosmic-ray generation and \textsc{Geant4} \cite{AGOSTINELLI2003250} for particle transport and detector modeling. The generated particle spectra and fluxes are consistent with established simulations (e.g., CORSIKA \citep{heck1998corsika, Samalan_2022}) and reproduce sea-level measurements. The detector geometry and materials implemented in \textsc{Geant4} match the experimental setup.

The information required for analysis, such as particle position, momentum, particle identification (PID), energy deposition, and other attributes, is recorded as particles traverse the signal-sensitive region of the RPC detector. This data is subsequently used for screening complete signals, calculating scattering angles, and reconstructing scattering points via the Point of Closest Approach (PoCA) method \cite{BONOMI2020103768}. The simulation accounts for the physical interactions of particles with materials, such as air or other metal materials, considering processes such as multiple scattering, ionization, bremsstrahlung, pair production, Coulomb scattering, and nuclear capture, which are tracked and explicitly marked for further analysis.

Using the simulation framework described above, a muon tomography system with identical dimensions, materials, and configurations to the experimental setup was simulated. The simulation included cosmic-ray scattering in air and lead, accounting for muons, electrons, photons, neutrons, pions, protons and their antiparticles. After incorporating the detector response, muons contribute to the total rate by about 84.2\%, while electrons and positrons by about 12.1\%. This is also in good agreement with previous studies \cite{Abbrescia2023}.

\section{Result and Analysis}

\begin{figure}
    \centering
    \includegraphics[width=0.56\columnwidth]{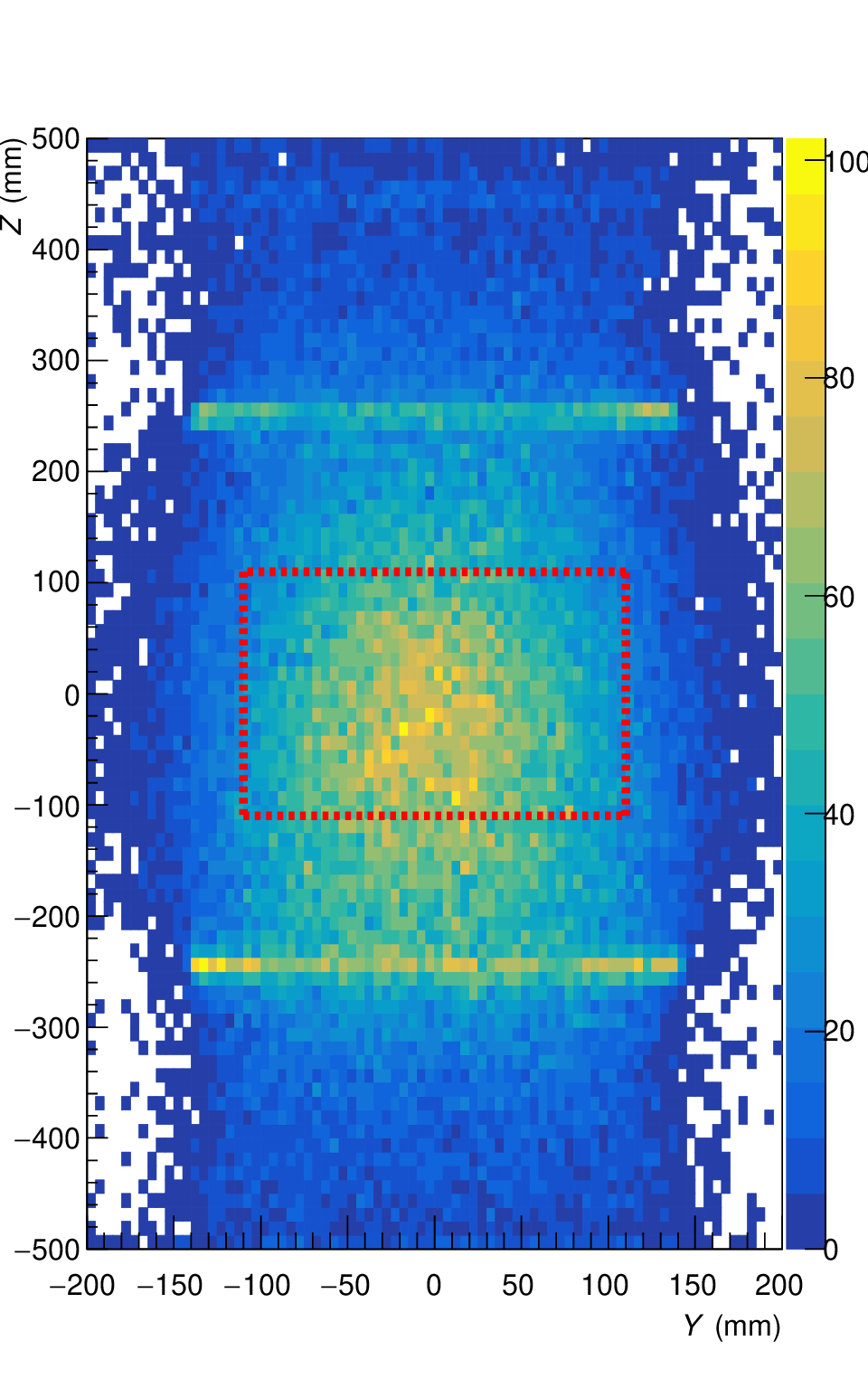}
    \captionsetup{justification=raggedright} % 局部左对齐
    \caption{PoCA position distribution in the $Y$-$Z$ plane. Scattering events with $\theta >$ 0.05 rad from the 63-day Physics RUN test. The fiducial volume, delineated by a red dashed contour, is defined within the coordinate range: $X$, $Y$, $Z$ $\in$ (-110 mm, 110 mm).}
    \label{fig:PoCA}
\end{figure}

By tracking the muon positions across four RPC layers, its scattering angle $\theta$ can be determined if deflection occurs between the second and third layers. The PoCA algorithm \cite{BONOMI2020103768} is employed, which approximates multiple scatterings as a single effective interaction within the material (e.g., air). As shown in Fig.~\ref{fig:PoCA}, a substantial number of large-angle scattering events occur on the RPC surfaces. These events are likely caused by interactions with construction materials (e.g., concrete), as detailed in Ref.~\cite{zhang2025revealingsecondaryparticlesignatures}. (In addition, the presence of construction materials reduces the fraction of electrons in cosmic rays \cite{1481535}.) However, these events may introduce significant bias in the search for DM and must therefore be excluded. To suppress such contamination, the analysis is restricted to events with scattering vertices located within the fiducial volume defined by $X$, $Y$, $Z$ $\in$ (-110 mm, 110 mm).

Initial simulations considered only muon scattering in air and lead but underestimated events above 0.05 rad, suggesting contributions from particles beyond muons. To resolve this, later simulations included the full spectrum of secondary cosmic rays—muons, electrons, photons, neutrons, pions, protons, and their antiparticles—consistent with the sensitivity of RPCs to non-muon particles \cite{Hadjiiska_2021}. For both Physics Run and Control Run configurations, $10^8$ particles were generated, yielding approximately one million effective events per case. An effective event is defined as one where all four RPC layers record hits within their sensitive regions. Hit positions are reconstructed from energy deposition, with $X$ and $Y$ determined by energy-weighted coordinates and smeared with Gaussian noise to match the measured 0.7 mm resolution, while $Z$ positions are fixed at 450 mm, 250 mm, –250 mm, and –450 mm \cite{zhang2025revealingsecondaryparticlesignatures}.

In this study, the analysis is restricted to the angular range of 0.05 to 0.5 rad. Scattering angles below 0.05 rad are excluded because they are strongly affected by detector angular resolution and dominated by multiple scattering in the materials, making it difficult to extract reliable information about the scattering process. This region is also primarily governed by Coulomb scattering rather than potential DM interactions. Angles above 0.5 rad are excluded due to limited statistics. Only muons and electrons are considered in the analysis, as they are the dominant components in the simulation. Other particles are grouped together to reduce statistical fluctuations caused by their low abundance. Although antiparticles differ from their corresponding particles in the low-energy region, the detector’s energy threshold suppresses these contributions, leading to nearly identical detectable energy spectra and very similar scattering angle distributions. Therefore, particles and antiparticles are analyzed collectively. The excluded angular regions and minor particle species will be investigated in future analyses as improved statistics and refined methods become available.

\begin{figure}
    \centering
    \includegraphics[width=1.\columnwidth]{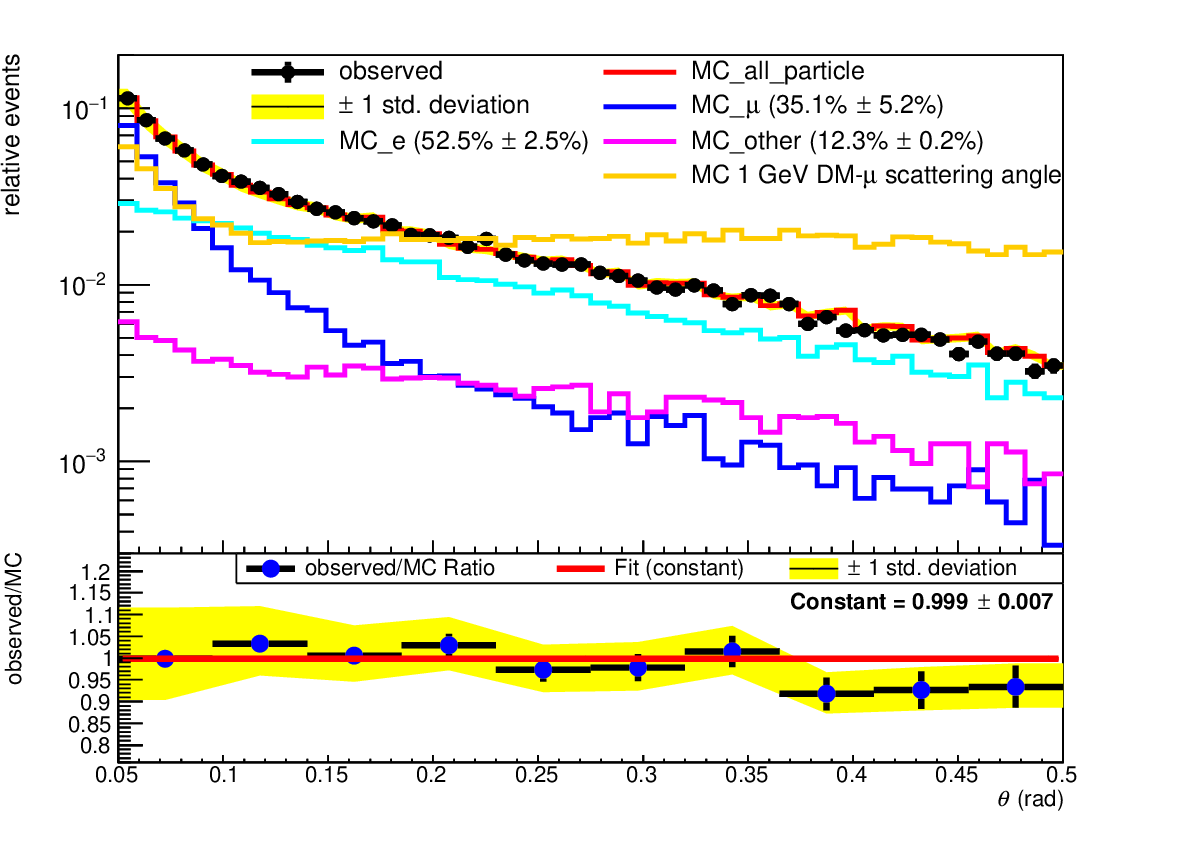}
    \captionsetup{justification=raggedright} % 局部左对齐
    \caption{Fitted $\theta$ distribution of Physics Run: data (black), total MC (red), muons (blue), electrons (cyan), others (magenta). Yellow bands indicate 1$\sigma$ regions. The orange line shows the simulated angular distribution for 1 GeV DM–muon scattering, with the cross section set to approximately $10^{-3}$ $\rm{cm}^2$. Lower panel shows data/MC ratios with a constant fit (red line).}
    \label{fig:fitted}
\end{figure}

Achieving higher precision in DM searches with cosmic-ray muons requires quantifying background composition. To determine sea-level particle fractions, a combined fit is applied using \textsc{RooFit} \cite{refId0} to match simulated scattering angle distributions of individual species to data. Only muon and electron fractions are varied, while other components are fixed due to their small contributions and negligible impact on the results. It is confirmed that fixing these minor components yielded results consistent with those obtained when they are allowed to vary. Using this fitting method on Physics Run, the ratio between observed and MC simulated distributions reaches $0.999 \pm 0.007$. The resulting particle composition is $(35.1 \pm 5.2)$\% muons and $(52.5 \pm 2.5)$\% electrons, as shown in Fig.~\ref{fig:fitted}. (At small scattering angles, the muon fraction becomes higher than that of electrons.)

Using the same simulation framework described earlier, the configuration of Control Run, as shown in Fig.\ref{fig:pb_experiment_setup}, is replicated to obtain simulated cosmic-ray data under identical detector conditions. This setup enables a clear division of the scattering region into a lead block zone, defined by scattering coordinates $X$ $\in$ (-110 mm, -50 mm), and an air zone, defined by $X$ $\in$ (50 mm, 110 mm). Applying the same fitting method, it is found that in the lead block region, muons account for $(96.6 \pm 0.2)\%$ and electrons for $(0.7 \pm 0.2)\%$. In the air region, the muon and electron fractions are $(32.3 \pm 1.4)\%$ and $(54.6 \pm 1.4)\%$, respectively. The low electron fraction in the lead block region is expected, as electrons are less penetrating than muons and cannot traverse the lead to reach the downstream RPC layers. The same fitting method is applied to a Physics Run, with results shown in Fig.~\ref{fig:fitted}. Restricting the scattering point range to the same $X$ region as the air zone in Control Run data yields a muon fraction of $(32.1 \pm 1.3)\%$ and an electron fraction of $(54.6 \pm 1.3)\%$, consistent with those observed in Control Run.

To evaluate the systematic uncertainty, the method's stability is tested by varying the angular range, scattering volume, and treatment of other particle components. This approach is analogous to transforming and calibrating a control region to constrain the distribution in the signal region \cite{Undagoitia_2015}. Re-fitting the angular subset in the range of 0.1–0.35 rad yields $f_1$, the ratio of particle fractions before and after the re-fitting. Similarly, selecting events within $X$, $Y$, $Z$ $\in$ (-100 mm, 100 mm) defines the ratio $f_2$, and allowing other components to vary defines $f_3$. The combined stability factor $F = {f}_{1}{f}_{2}{f}_{3}$ is used to quantify robustness. Along with the fit uncertainty ${\sigma}_{\rm{fit}}$,  the relative contribution of this method $\sigma_{\rm{all}}$ is given by Eq.~(\ref{eq:1}), where $C_{\rm{fit}}$ is the initially fitted particle fraction. The results are summarized in Table~\ref{tab:uncertainty}, where $C_{\rm{unfit}}$ is the particle fraction without any fitted. Based on this fit, the relative contributions of different particle species in secondary cosmic rays at sea level were extracted.

\begin{equation}
    \sigma_{\rm{all}} = \sqrt{{{\sigma}_{\rm{fit}}}^{2} + {(C_{\rm{fit}}|1-F|)}^{2}} .
    \label{eq:1}
\end{equation}

\begin{table}[htbp]
    \centering
    \begin{tabular}{c|c|c}
        \hline\hline
          $\rm{Particle Type}$                   & $\mu$     & $e$       \\ \hline
          $C_{\rm{unfit}}$                       & $32.2\%$ & $55.4\%$ \\ \hline
          $C_{\rm{fit}}$                         & $35.2\%$ & $52.5\%$ \\ \hline
          ${\sigma}_{\rm{fit}}$                  & $0.5\%$  & $0.5\%$  \\ \hline
          $f_1$                                  & $1.2$    & $1.0$    \\ \hline
          $f_2$                                  & $1.0$    & $1.0$    \\ \hline
          $f_3$                                  & $1.0$    & $1.1$    \\ \hline
          $F$                                    & $1.2$    & $1.0$    \\ \hline
          $\sigma_{\rm{all}}$                               & $5.2\%$  & $2.5\%$  \\ 
        \hline\hline
    \end{tabular}
    \caption{Systematic uncertainty components for sea-level secondary cosmic-ray muon and electron fractions}
    \label{tab:uncertainty}
\end{table}

Based on the cosmic-ray composition extracted from the measurements and the tomography system implemented in the simulation framework, the scattering angle distributions arising from interactions between sea-level cosmic-ray muons and DM particles of varying masses are further simulated. Fig.~\ref{fig:fitted} shows the scattering angle distribution for a 1 GeV DM-muon interaction. The velocity distribution of DM is assumed to be 220 $\rm{km/s}$, with a density of 0.3 $\rm{GeV/cm^3}$ \cite{PhysRevD.110.016017}, uniformly distributed in space. Model-independent elastic scattering between muons and DM is investigated under Newtonian mechanics. By applying the principles of energy and momentum conservation, the muon recoil energy $E_{\rm{recoil}}$ is derived as expressed in Ref.~\cite{PhysRevD.110.016017}. Therefore, compared to other experiments such as NA64$\mu$ \cite{PhysRevLett.132.211803}, our results exhibit a higher degree of independence from specific reaction models.

Traditional DM experiments, such as XENON1T \cite{Aprile2017} and PandaX \cite{PhysRevLett.119.181302}, treat atomic nuclei as quasi-static targets, assuming DM to be the incident particle. In contrast, in this experiment, the average energy of cosmic-ray muons at sea level is approximately 3-4 GeV, placing them in a relativistic state. Therefore, from the perspective of the high-speed incident muons, DM is essentially at rest.

The calculation of the muon–DM scattering cross section requires several physical inputs. These include the sea-level muon flux $F_{\mu} \approx 1/60~\rm{s^{-1}cm^{-2}}$, the DM mass $M_{\rm{DM}}$, and a detection efficiency of $\epsilon_1 = (53.68 \pm 2.88)\%$ for the full four-layer RPC system. In the subsequent analysis, the same selection criteria used in the fit are applied: scattering angles in the range of 0.05 to 0.5 rad and scattering positions within $X$, $Y$, $Z$ $\in$ (-110 mm, 110 mm). Therefore, $V$ is the sensitive volume, given as 22 cm $\times$ 22 cm $\times$ 22 cm $=$ 10,648 $\rm{cm^3}$.

To further suppress background, reconstructed scattering points are required to lie within the sensitive volume. Accordingly, $\Omega_a$ is defined as the angular acceptance probability for signal events under this geometry, with values for different $M_{\rm{DM}}$ assumptions listed in Table~\ref{tab:bkgnum_sigeff}. Here, the detection efficiency does not refer to the probability that all four detector layers are triggered. Instead, since $\Omega_a$ is defined as the fraction of signal events where the lower two layers are triggered, given that the upper two layers have already been triggered, the relevant detection efficiency, $\epsilon_2$, corresponds to the product of the efficiencies of the lower two layers. This yields $\epsilon_2$ = $(71.32 \pm 2.57)\%$.

\begin{table}[htbp]
    \centering
    \begin{tabular}{c|c}
        \hline\hline
          $M_{\rm{DM}}$ $\rm{(GeV)}$      & $\Omega_a$ (\%)   \\ \hline
          1 $\times$ $10^{-2}$            & 66.7 $\pm$ 1.3 \\ \hline
          5 $\times$ $10^{-2}$            & 67.1 $\pm$ 1.3 \\ \hline
          1 $\times$ $10^{-1}$            & 67.1 $\pm$ 1.3 \\ \hline
          5 $\times$ $10^{-1}$            & 69.2 $\pm$ 1.3 \\ \hline
          1 $\times$ $10^{0}$             & 67.2 $\pm$ 1.3 \\ \hline
          5 $\times$ $10^{0}$             & 63.8 $\pm$ 1.3 \\ \hline
          1 $\times$ $10^{1}$             & 63.8 $\pm$ 1.2 \\ \hline
          5 $\times$ $10^{1}$             & 63.8 $\pm$ 1.2 \\ \hline
          1 $\times$ $10^{2}$             & 62.7 $\pm$ 1.2 \\ 
        \hline\hline
    \end{tabular}
    \caption{Signal angular acceptance probabilities for different ${\rm{M}}_{\rm{DM}}$ assumptions.}
    \label{tab:bkgnum_sigeff}
\end{table}

Eq.~(\ref{eq:6}) follows from the above considerations. 
\begin{equation}
    dN/dt=\rho V/{M}_{\rm DM} \times \sigma_{\mu,\rm{DM}} \times F_{\mu} \times \epsilon_1 \times \Omega_a \times \epsilon_2 .
    \label{eq:6}
\end{equation}
However, if a strongly interacting dark matter component is captured and thermalized within the Earth, its local surface density can be enormously enhanced for DM with a mass of 1 GeV, the density may increase by a factor of $10^{15}$ \cite{PhysRevLett.131.011005, PhysRevD.109.075027}. To account for this possibility, an enhancement factor $f_{E} = 10^{15}$ is introduced, such that $D_{\rm{E}} = f_{E} \rho / D_{\rm{DM}}$, where $D_{\rm{E}}$ denotes the number density of the exotic slow DM under investigation, and $D_{\rm{DM}}$ is the $M_{\rm{DM}}$ in unit GeV.

\begin{figure}[htbp]
    \centering
    \includegraphics[width=\linewidth]{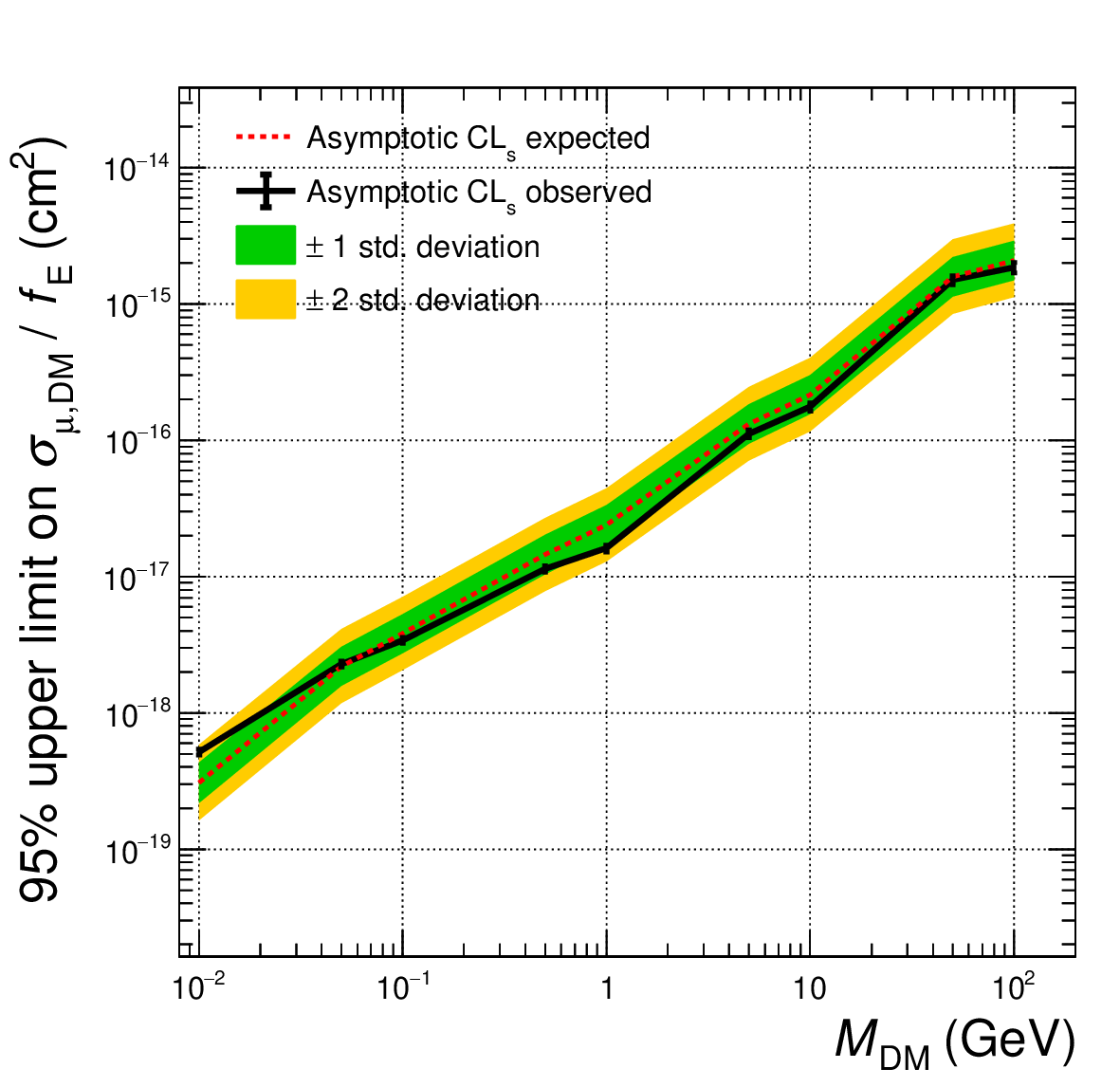}%
    \captionsetup{justification=raggedright} % 局部左对齐
    \caption{Expected and observed 95\% CL upper limits on the DM-muon interaction cross section versus ${M}_{\rm{DM}}$, assuming a local DM density enhancement factor of $10^{15}$ arising from the Earth-bound thermalized component. The green and yellow bands represent the 1$\sigma$ and 2$\sigma$ regions, respectively.}
    \label{fig:sigmalimit} 
\end{figure}

A binned maximum likelihood fit is performed on the observed, background, and signal samples using the HiggsCombine statistical analysis tool \cite{CMS:2024onh}. The $\rm{CL_s}$ method is applied to set the upper limit (UL) \citep{JUNK1999435,ALRead_2002}. Since the total number of events measured in the experiment is given by $N_{\rm{all}} = F_{\mu} \times \epsilon_1 \times t \times A$, where $A$ = 22 cm $\times$ 22 cm = 484 $\rm{cm^2}$, Eq.~(\ref{eq:6}) can be extended to as: 
\begin{equation}
    {\sigma}_{\mu, \rm{DM}} = \frac{N{M}_{\rm DM}A}{\rho VN_{\rm{all}}\Omega_a\epsilon_2f_{E}}.
    \label{eq:7}
\end{equation}
Using Eq.~(\ref{eq:7}), the DM–muon elastic scattering cross section is calculated, and the results are shown in Fig.~\ref{fig:sigmalimit}. In the uncertainty estimation, both the systematic uncertainty from the fitting method and the statistical uncertainties associated with each term in Eq~(\ref{eq:7}), are taken into account. The figure displays the expected and observed 95\% confidence level (CL) upper limits on the DM–muon scattering cross section, based on the cosmic-ray scattering data collected under laboratory air conditions. For a 1 GeV DM particle, the upper limit on the scattering cross section is approximately $1.62 \times 10^{-17}$ $\rm{cm^{2}}$.

\section{Summary and Outlook}
In this work, an RPC-based muon tomography system was deployed for high-precision angular scattering measurements of cosmic-ray muons at sea level. Combining detailed simulations and a multi-template fitting procedure, this work resolved the relative abundances of secondary cosmic-ray muons and electrons with good precision -- reaching $\sim 2\%$ for the electron component. This resolution mitigates long-standing uncertainties in sea-level particle composition and establishes the muon tomography system as a powerful tool for cosmic-ray spectroscopy.

Furthermore, we pioneered a direct probe for light muon-philic dark matter via cosmic-ray muon–DM elastic scattering. Applying a model-independent analysis framework to 63-day Physics Run data, this work derived constraints on the slow DM–muon cross-section across the 0.1–10 GeV mass range. For 1 GeV DM, the 95\% CL limit reaches $1.61 \times 10^{-17}$ $\rm{cm}^{2}$, demonstrating sensitivity to sub-GeV DM couplings. Future operation with higher statistics is expected to improve the determination of secondary cosmic-ray composition, enabling sensitivity to additional particle species. In parallel, to enhance the sensitivity of DM searches, efforts will focus on experiments utilizing high-intensity, well-collimated muon beams, along with extended cosmic-ray runs employing larger detectors and longer exposure times. With an enlarged detector volume of up to 1 $\rm{m^3}$, an extended operation period of approximately one year, and by further incorporating the full scattering angle range, the overall sensitivity is expected to improve by nearly 4-5 orders of magnitude. Such an improvement would allow this model-independent approach to probe unexplored regions of parameter space in muon-philic dark matter scenarios and potentially surpass existing limits from beam experiments in the GeV mass range and beyond.

\section{Acknowledgments}
This work is supported in part by the National Natural Science Foundation of China under Grants No. 12325504. This work is also supported in part by the State Key Laboratory of Nuclear Physics and Technology, Peking University (No. NPT2022ZZ02). 

The data acquisition system in this work is supported by technical assistance from Zhongtao Shen and Hao Liu at the University of Science and Technology of China. We sincerely thank them for their support.

% The \nocite command causes all entries in a bibliography to be printed out
% whether or not they are actually referenced in the text. This is appropriate
% for the sample file to show the different styles of references, but authors
% most likely will not want to use it.
\nocite{*}

\bibliography{apssamp}% Produces the bibliography via BibTeX.

@article{PhysRevD.110.016017,
  title = {Proposed Peking University muon experiment for muon tomography and dark matter search},
  author = {Yu, Xudong and Wang, Zijian and Liu, Cheng-en and others},
  journal = {Phys. Rev. D},
  volume = {110},
  issue = {1},
  pages = {016017},
  numpages = {14},
  year = {2024},
  month = {Jul},
  publisher = {American Physical Society},
  doi = {10.1103/PhysRevD.110.016017},
  url = {https://link.aps.org/doi/10.1103/PhysRevD.110.016017}
}

@article{BONOMI2020103768,
title = {Applications of cosmic-ray muons},
journal = {Progress in Particle and Nuclear Physics},
volume = {112},
pages = {103768},
year = {2020},
issn = {0146-6410},
doi = {https://doi.org/10.1016/j.ppnp.2020.103768},
url = {https://www.sciencedirect.com/science/article/pii/S0146641020300156},
author = {G. Bonomi and P. Checchia and M. D’Errico and others}
}

@Article{Aprile2017,
author={Aprile, E.
and Aalbers, J.
and Agostini, F.
and others},
title={The XENON1T dark matter experiment},
journal={The European Physical Journal C},
year={2017},
month={Dec},
day={18},
volume={77},
number={12},
pages={881},
issn={1434-6052},
doi={10.1140/epjc/s10052-017-5326-3},
url={https://doi.org/10.1140/epjc/s10052-017-5326-3}
}

@article{PhysRevLett.119.181302,
  title = {Dark Matter Results from 54-Ton-Day Exposure of PandaX-II Experiment},
  author = {Cui, Xiangyi and Abdukerim, Abdusalam and Chen, Wei and others},
  collaboration = {PandaX-II Collaboration},
  journal = {Phys. Rev. Lett.},
  volume = {119},
  issue = {18},
  pages = {181302},
  numpages = {7},
  year = {2017},
  month = {Oct},
  publisher = {American Physical Society},
  doi = {10.1103/PhysRevLett.119.181302},
  url = {https://link.aps.org/doi/10.1103/PhysRevLett.119.181302}
}

@article{JUNK1999435,
title = {Confidence level computation for combining searches with small statistics},
journal = {Nuclear Instruments and Methods in Physics Research Section A: Accelerators, Spectrometers, Detectors and Associated Equipment},
volume = {434},
number = {2},
pages = {435-443},
year = {1999},
issn = {0168-9002},
doi = {https://doi.org/10.1016/S0168-9002(99)00498-2},
url = {https://www.sciencedirect.com/science/article/pii/S0168900299004982},
author = {Thomas Junk}
}

@article{ALRead_2002,
doi = {10.1088/0954-3899/28/10/313},
url = {https://dx.doi.org/10.1088/0954-3899/28/10/313},
year = {2002},
month = {sep},
publisher = {},
volume = {28},
number = {10},
pages = {2693},
author = {A L Read},
title = {Presentation of search results: the CLs technique},
journal = {Journal of Physics G: Nuclear and Particle Physics}
}

@book{rao1998extensive,
  title={Extensive air showers},
  author={Rao, Mangu VS and Sreekantan, Badanaval Venkata},
  year={1998},
  publisher={World scientific}
}

@article{Workman:2022ynf,
    author = "Workman, R. L. and others",
    collaboration = "Particle Data Group",
    title = "{Review of Particle Physics}",
    doi = "10.1093/ptep/ptac097",
    journal = "PTEP",
    volume = "2022",
    pages = "083C01",
    year = "2022"
}

@article{doi:10.1139/p68-411,
    author = {Beuermann, K. P. and Wibberenz, G.},
    title = {Secondary spectra of electrons and photons in the atmosphere},
    journal = {Canadian Journal of Physics},
    volume = {46},
    number = {10},
    pages = {S1034-S1037},
    year = {1968},
    doi = {10.1139/p68-411},
    URL = {https://doi.org/10.1139/p68-411}
}

@book{grieder2001cosmic,
  title={Cosmic rays at Earth},
  author={Grieder, Peter KF},
  year={2001},
  publisher={Elsevier}
}

@book{miroshnichenko2015solar,
  title={Solar Cosmic rays: fundamentals and applications},
  author={Miroshnichenko, Leonty},
  year={2015},
  publisher={Springer}
}

@article{sikdar_2023,
    title={Study of secondary cosmic rays using small stratospheric balloon missions},
    author={Rupnath Sikdar and Sandip K. Chakrabarti and Debashis Bhowmick},
    volume={44},
    url={http://dx.doi.org/10.1007/s12036-023-09964-6},
    doi={10.1007/s12036-023-09964-6},
    number={2},
    journal={Journal of Astrophysics and Astronomy},
    publisher={Springer Science and Business Media LLC},
    year={2023},
    month={Jul},
    language={en},
}

@misc{essig2023snowmass2021cosmicfrontierlandscape, 
      author={Rouven Essig and Graham K. Giovanetti and Noah Kurinsky and others},
      year={2023},
      eprint={2203.08297},
      archivePrefix={arXiv},
      primaryClass={hep-ph},
      url={https://arxiv.org/abs/2203.08297}, 
}

@misc{harris2022snowmasswhitepapernew,
      author={Philip Harris and Philip Schuster and Jure Zupan},
      year={2022},
      eprint={2207.08990},
      archivePrefix={arXiv},
      primaryClass={hep-ph},
      url={https://arxiv.org/abs/2207.08990}, 
}

@article{Bai2014,
  author    = {Bai, Yang and Berger, Joshua},
  title     = {Lepton Portal dark matter},
  journal   = {Journal of High Energy Physics},
  year      = {2014},
  volume    = {2014},
  number    = {8},
  pages     = {153},
  doi       = {10.1007/JHEP08(2014)153},
  url       = {https://doi.org/10.1007/JHEP08(2014)153},
  issn      = {1029-8479},
}

@article{annurev:/content/journals/10.1146/annurev.ns.33.120183.002123,
   author = "Drees, J and Montgomery, H E",
   title = "Muon Scattering", 
   journal= "Annual Review of Nuclear and Particle Science",
   year = "1983",
   volume = "33",
   number = "Volume 33, 1983",
   pages = "383-452",
   doi = "https://doi.org/10.1146/annurev.ns.33.120183.002123",
   url = "https://www.annualreviews.org/content/journals/10.1146/annurev.ns.33.120183.002123",
   publisher = "Annual Reviews",
   issn = "1545-4134",
   type = "Journal Article",
  }

@article{PhysRevLett.132.211803,
  title = {First Results in the Search for Dark Sectors at NA64 with the CERN SPS High Energy Muon Beam},
  author = {Andreev, Yu. M. and Banerjee, D. and Banto Oberhauser, B. and others},
  journal = {Phys. Rev. Lett.},
  volume = {132},
  issue = {21},
  pages = {211803},
  numpages = {7},
  year = {2024},
  month = {May},
  publisher = {American Physical Society},
  doi = {10.1103/PhysRevLett.132.211803},
  url = {https://link.aps.org/doi/10.1103/PhysRevLett.132.211803}
}

@article{CARLONICALAME2015325,
title = {A new approach to evaluate the leading hadronic corrections to the muon g-2},
journal = {Physics Letters B},
volume = {746},
pages = {325-329},
year = {2015},
issn = {0370-2693},
doi = {https://doi.org/10.1016/j.physletb.2015.05.020},
url = {https://www.sciencedirect.com/science/article/pii/S0370269315003573},
author = {C.M. {Carloni Calame} and M. Passera and L. Trentadue and others}
}

@article{doi:10.1142/S0217732325300083,
author = {Gao, Leyun and Liu, Cheng-En and Li, Qite and Zhou, Chen and Li, Qiang and Chen, Liangwen and Zhang, Xueheng and Xu, Yu and Sun, Zhiyu},
title = {Probing and knocking with muons},
journal = {Modern Physics Letters A},
volume = {40},
number = {24},
pages = {2530008},
year = {2025},
doi = {10.1142/S0217732325300083},
URL = { https://doi.org/10.1142/S0217732325300083}
}

@article{PhysRevLett.131.011005,
  title = {Dark Matter Annihilation inside Large-Volume Neutrino Detectors},
  author = {McKeen, David and Morrissey, David E. and Pospelov, Maxim and others},
  journal = {Phys. Rev. Lett.},
  volume = {131},
  issue = {1},
  pages = {011005},
  numpages = {7},
  year = {2023},
  month = {Jul},
  publisher = {American Physical Society},
  doi = {10.1103/PhysRevLett.131.011005},
  url = {https://link.aps.org/doi/10.1103/PhysRevLett.131.011005}
}

@article{PhysRevD.109.075027,
  title = {Terrestrial density of strongly-coupled relics},
  author = {Berlin, Asher and Liu, Hangwan and Pospelov, Maxim and others},
  journal = {Phys. Rev. D},
  volume = {109},
  issue = {7},
  pages = {075027},
  numpages = {25},
  year = {2024},
  month = {Apr},
  publisher = {American Physical Society},
  doi = {10.1103/PhysRevD.109.075027},
  url = {https://link.aps.org/doi/10.1103/PhysRevD.109.075027}
}

@article{PhysRevD.103.115031,
  title = {Earth-bound millicharge relics},
  author = {Pospelov, Maxim and Ramani, Harikrishnan},
  journal = {Phys. Rev. D},
  volume = {103},
  issue = {11},
  pages = {115031},
  numpages = {16},
  year = {2021},
  month = {Jun},
  publisher = {American Physical Society},
  doi = {10.1103/PhysRevD.103.115031},
  url = {https://link.aps.org/doi/10.1103/PhysRevD.103.115031}
}

@article{LI201222,
title = {Study of spatial resolution properties of a glass RPC},
journal = {Nuclear Instruments and Methods in Physics Research Section A: Accelerators, Spectrometers, Detectors and Associated Equipment},
volume = {663},
number = {1},
pages = {22-25},
year = {2012},
issn = {0168-9002},
doi = {https://doi.org/10.1016/j.nima.2011.10.006},
url = {https://www.sciencedirect.com/science/article/pii/S0168900211018912},
author = {Li, Qite and Ye, Yanlin and Wen, Chao and others}
}

@article{Li_2013,
doi = {10.1088/1674-1137/37/1/016002},
url = {https://dx.doi.org/10.1088/1674-1137/37/1/016002},
year = {2013},
month = {jan},
publisher = {},
volume = {37},
number = {1},
pages = {016002},
author = {Li, Qite and Ye, Yanlin and Ji, Wei and others},
title = {A sub-millimeter spatial resolution achieved by a large sized glass RPC},
journal = {Chinese Physics C}
}

@article{https://doi.org/10.1029/92JA02672,
author = {De Pascale, M. P. and Morselli, A. and Picozza, P. and others},
title = {Absolute spectrum and charge ratio of cosmic ray muons in the energy region from 0.2 GeV to 100 GeV at 600 m above sea level},
journal = {Journal of Geophysical Research: Space Physics},
volume = {98},
number = {A3},
pages = {3501-3507},
doi = {https://doi.org/10.1029/92JA02672},
url = {https://agupubs.onlinelibrary.wiley.com/doi/abs/10.1029/92JA02672},
year = {1993}
}

@article{PhysRevLett.83.4241,
  title = {Measurements of Ground-Level Muons at Two Geomagnetic Locations},
  author = {Kremer, J. and Boezio, M. and Ambriola, M. L. and others},
  journal = {Phys. Rev. Lett.},
  volume = {83},
  issue = {21},
  pages = {4241--4244},
  numpages = {0},
  year = {1999},
  month = {Nov},
  publisher = {American Physical Society},
  doi = {10.1103/PhysRevLett.83.4241},
  url = {https://link.aps.org/doi/10.1103/PhysRevLett.83.4241}
}

@article{HAINO200435,
title = {Measurements of primary and atmospheric cosmic-ray spectra with the BESS-TeV spectrometer},
journal = {Physics Letters B},
volume = {594},
number = {1},
pages = {35-46},
year = {2004},
issn = {0370-2693},
doi = {https://doi.org/10.1016/j.physletb.2004.05.019},
url = {https://www.sciencedirect.com/science/article/pii/S0370269304007567},
author = {S. Haino and T. Sanuki and K. Abe and others}
}

@article{ACHARD200415,
title = {Measurement of the atmospheric muon spectrum from 20 to 3000 GeV},
journal = {Physics Letters B},
volume = {598},
number = {1},
pages = {15-32},
year = {2004},
issn = {0370-2693},
doi = {https://doi.org/10.1016/j.physletb.2004.08.003},
url = {https://www.sciencedirect.com/science/article/pii/S0370269304011438},
author = {P. Achard and O. Adriani and M. Aguilar-Benitez and others}
}

@article{ALLKOFER1971425,
title = {The absolute cosmic ray muon spectrum at sea level},
journal = {Physics Letters B},
volume = {36},
number = {4},
pages = {425-427},
year = {1971},
issn = {0370-2693},
doi = {https://doi.org/10.1016/0370-2693(71)90741-6},
url = {https://www.sciencedirect.com/science/article/pii/0370269371907416},
author = {O.C. Allkofer and K. Carstensen and W.D. Dau}
}

@article{B_C_Rastin_1984,
doi = {10.1088/0305-4616/10/11/017},
url = {https://dx.doi.org/10.1088/0305-4616/10/11/017},
year = {1984},
month = {nov},
publisher = {},
volume = {10},
number = {11},
pages = {1609},
author = {B C Rastin},
title = {An accurate measurement of the sea-level muon spectrum within the range 4 to 3000 GeV/c},
journal = {Journal of Physics G: Nuclear Physics}
}

@article{C_A_Ayre_1975,
doi = {10.1088/0305-4616/1/5/010},
url = {https://dx.doi.org/10.1088/0305-4616/1/5/010},
year = {1975},
month = {jun},
publisher = {},
volume = {1},
number = {5},
pages = {584},
author = {C A Ayre and J M Baxendale and C J Hume and others},
title = {Precise measurement of the vertical muon spectrum in the range 20-500 GeV/c},
journal = {Journal of Physics G: Nuclear Physics}
}

@article{PhysRevD.19.1368,
  title = {Cosmic-ray muon spectrum up to 1 TeV at 75\ifmmode^\circ\else\textdegree\fi{} zenith angle},
  author = {Jokisch, H. and Carstensen, K. and Dau, W. D. and others},
  journal = {Phys. Rev. D},
  volume = {19},
  issue = {5},
  pages = {1368--1372},
  numpages = {0},
  year = {1979},
  month = {Mar},
  publisher = {American Physical Society},
  doi = {10.1103/PhysRevD.19.1368},
  url = {https://link.aps.org/doi/10.1103/PhysRevD.19.1368}
}

@article{Allkofer1970,
  author    = {Allkofer, O. C. and Knoblich, P.},
  title     = {The momentum spectra of cosmic-ray particles at sea level in the momentum range (0.05÷6) GeV/c},
  journal   = {Lettere al Nuovo Cimento (1969-1970)},
  year      = {1970},
  volume    = {3},
  number    = {1},
  pages     = {6--8},
  doi       = {10.1007/BF02755767},
  url       = {https://doi.org/10.1007/BF02755767},
  issn      = {1827-613X},
}

@article{AGOSTINELLI2003250,
title = {Geant4—a simulation toolkit},
journal = {Nuclear Instruments and Methods in Physics Research Section A: Accelerators, Spectrometers, Detectors and Associated Equipment},
volume = {506},
number = {3},
pages = {250-303},
year = {2003},
issn = {0168-9002},
doi = {https://doi.org/10.1016/S0168-9002(03)01368-8},
url = {https://www.sciencedirect.com/science/article/pii/S0168900203013688},
author = {S. Agostinelli and J. Allison and K. Amako and others}
}

@INPROCEEDINGS{4437209,
  author={Hagmann, Chris and Lange, David and Wright, Douglas},
  booktitle={2007 IEEE Nuclear Science Symposium Conference Record}, 
  title={Cosmic-ray shower generator (CRY) for Monte Carlo transport codes}, 
  year={2007},
  volume={2},
  number={},
  pages={1143-1146},
  doi={10.1109/NSSMIC.2007.4437209}}

@article{ refId0,
	author = {{Verkerke, Wouter}},
	title = {RooFit},
	DOI= "10.1051/epjconf/20100402005",
	url= "https://doi.org/10.1051/epjconf/20100402005",
	journal = {EPJ Web of
                    Conferences},
	year = 2010,
	volume = 4,
	pages = "02005",
}

@article{heck1998corsika,
  title={CORSIKA: A Monte Carlo code to simulate extensive air showers},
  author={Heck, Dieter and Knapp, Johannes and Capdevielle, JN and others},
  journal={Report fzka},
  volume={6019},
  number={11},
  year={1998}
}

@article{Samalan_2022,
doi = {10.1088/1748-0221/17/01/C01015},
url = {https://dx.doi.org/10.1088/1748-0221/17/01/C01015},
year = {2022},
month = {jan},
publisher = {IOP Publishing},
volume = {17},
number = {01},
pages = {C01015},
author = {Samalan, A. and Basnet, S. and Bonechi, L. and others},
title = {End-to-end simulations of the MUon RAdiography of VESuvius experiment},
journal = {Journal of Instrumentation}
}

@article{
    CMS:2024onh,
    author = "Hayrapetyan, Aram and others",
    collaboration = "CMS",
    title = "The {CMS} statistical analysis and combination tool: {\textsc{Combine}}",
    eprint = "2404.06614",
    archivePrefix = "arXiv",
    primaryClass = "physics.data-an",
    reportNumber = "CMS-CAT-23-001, CERN-EP-2024-078",
    year = "2024",
    journal = "Comput. Softw. Big Sci.",
    doi = "10.1007/s41781-024-00121-4",
    volume = "8",
    pages = "19"
}

@article{Undagoitia_2015,
   title={Dark matter direct-detection experiments},
   volume={43},
   ISSN={1361-6471},
   url={http://dx.doi.org/10.1088/0954-3899/43/1/013001},
   DOI={10.1088/0954-3899/43/1/013001},
   number={1},
   journal={Journal of Physics G: Nuclear and Particle Physics},
   publisher={IOP Publishing},
   author={Undagoitia, Teresa Marrodán and Rauch, Ludwig},
   year={2015},
   month=dec, pages={013001} }

@misc{zhang2025revealingsecondaryparticlesignatures,
      author={Zhang, Rongfeng and Qin, Zibo and Liu, Cheng-en and others},
      year={2025},
      eprint={2507.03914},
      archivePrefix={arXiv},
      primaryClass={hep-ex},
      url={https://arxiv.org/abs/2507.03914}, 
}

@article{Hadjiiska_2021,
doi = {10.1088/1748-0221/16/04/C04005},
url = {https://dx.doi.org/10.1088/1748-0221/16/04/C04005},
year = {2021},
month = {apr},
publisher = {IOP Publishing},
volume = {16},
number = {04},
pages = {C04005},
author = {Hadjiiska, R. and Samalan, A. and others},
collaboration = "CMS",
title = {CMS RPC background — studies and measurements},
journal = {Journal of Instrumentation}
}

@misc{Liu_PKMuDM_2025,
  author       = {PKMu Collaboration},
  title        = {PKMu\_DM: Observed data and analysis code},
  howpublished = {GitHub repository, \url{https://github.com/Cheng-enLiu/PKMu_DM}},
  year         = {2025}
}

@misc{MC_construction,
  author       = {PKMu Collaboration},
  title        = {PKMUON\_2024: newrpc.yaml configuration file},
  howpublished = {GitHub repository, \url{https://github.com/PKMuon/PKMUON_2024}},
  year         = {2024}
}

@article{ParticleDataGroup:2024cfk,
    author = "Navas, S. and others",
    collaboration = "Particle Data Group",
    title = "{Review of particle physics}",
    doi = "10.1103/PhysRevD.110.030001",
    journal = "Phys. Rev. D",
    volume = "110",
    number = "3",
    pages = "030001",
    year = "2024"
}

@article{Abbrescia2023,
  author = {M. Abbrescia and C. Avanzini and L. Baldini and others},
  title = {Measurement of the cosmic charged particle rate at sea level in the latitude range 35$^{\circ}$ {\textendash} $82^{\circ}$N with the PolarquEEEst experiment},
  journal = {The European Physical Journal C},
  year = {2023},
  volume = {83},
  number = {4},
  pages = {293},
  doi = {10.1140/epjc/s10052-023-11353-w},
  url = {https://doi.org/10.1140/epjc/s10052-023-11353-w},
  issn = {1434-6052}
}

@ARTICLE{1481535,
  author={Ziegler, J.F.},
  journal={IEEE Transactions on Electron Devices}, 
  title={The effect of concrete shielding on cosmic ray induced soft fails in electronic systems}, 
  year={1981},
  volume={28},
  number={5},
  pages={560-565},
  doi={10.1109/T-ED.1981.20383}}
\bibliographystyle{apsrev4-2}

\end{document}